\documentclass[english,aps,prb,twocolumn,superscriptaddress,amsmath,amssymb,showpacs,showkeys]{revtex4}
\usepackage[T1]{fontenc}
\usepackage[latin1]{inputenc}
\usepackage{graphicx}
\usepackage{subfigure}

\makeatletter

\providecommand{\tabularnewline}{\\}

\usepackage{epsfig}

\usepackage{prettyref}

\usepackage{babel}
\makeatother

\begin{document}


\title{Slow Dynamics in Hard Condensed Matter: A Case Study of the Phase Separating System NdNiO$_3$}

\author{Devendra Kumar}

\affiliation{Department of Physics, Indian Institute of Technology Kanpur 208016,
India}

\author{K.P. Rajeev}

\email{kpraj@iitk.ac.in}

\affiliation{Department of Physics, Indian Institute of Technology Kanpur 208016,
India}

\author{J. A. Alonso}

\affiliation{Instituto de Ciencia de Materiales de Madrid, CSIC, Cantoblanco,
E-28049 Madrid, Spain}

\author{M. J. Martínez-Lope}

\affiliation{Instituto de Ciencia de Materiales de Madrid, CSIC, Cantoblanco,
E-28049 Madrid, Spain}

\begin{abstract}
We report the time dependent response of electrical resistivity in the non-magnetic perovskite oxide NdNiO$_3$ in its phase separated state and provide a physical explanation of the observations. We also model the system and do an accurate Monte Carlo simulation of the observed behavior. While cooling a phase separation takes place in this system below its metal-insulator transition temperature and in this state the material exhibits various dynamical phenomena such as relaxation of resistivity, dependence of resistivity on cooling rate and rejuvenation of the material after ageing. These phenomena signal that the phase separated state of NdNiO$_3$ is not in thermodynamic equilibrium and we conjecture that it consists of supercooled paramagnetic metallic and antiferromagnetic insulating phases. The supercooled phases are metastable and they switch over to the insulating equilibrium state stochastically and this  can account for the slow dynamics observed in our system. We also verify the predictive power of our model by simulating the result of a new experiment and confirming it by actual measurements.
\end{abstract}

\pacs{64.60.My, 64.70.K-, 71.30.+h,}

\keywords{Phase Separation, Hysteresis, Supercooling, Relaxation, Time dependence, Metal-Insulator Transition}

\maketitle

\section{INTRODUCTION}
Usually hard condensed matter is associated with fast dynamics, but sometimes we also see slow dynamics, particularly in cases such as glasses, spin glasses, phase separated systems etc. Dynamics of glasses and spin glasses have been intensively studied from the middle of last century, but such studies on phase separated systems started attracting a good deal of interest from physicists only towards the end of last century. Phase separation always seems to be associated with a broadened first order phase transition and it has been observed in such systems as transition metal oxides, a famous example being the colossal magnetoresistance (CMR) manganites, and alloys which show martensitic phase transitions.\cite{Zhang,Qazilbash,Chaddah,Dagotto,Lorenz}

There has been a large number of studies on the phase separated states associated with a broad first order metal-insulator (M-I) transition in transition metal oxides such as  VO$_2$, CMR manganites, the rare earth nickelate PrNiO$_3$, the cobaltite La$_{1-x}$Sr$_x$CoO$_3$ etc. They exhibit unusual phenomena such as very slow relaxation of resistivity\cite{Morin,Chen,Levy,Granados_1,J.Wu, Chaddah11} and magnetization,\cite{Ghivelder,Rivadulla} rejuvenation of resistivity after ageing and dependence of physical properties on cooling rate.\cite{Levy, Fischer, Chaddah10,Uehara}  A significant amount of work has gone in to study these systems in detail but they are not, as yet, well understood even though there is a recognition that the phenomena described above originate from phase separation.   The presence of the slow glass-like relaxations in the above systems has led many people to consider the phase separated systems as spin glasses, cluster glasses and it has been argued that the intercluster magnetic interactions are responsible for the observed glass-like dynamics in these systems.\cite{Dagotto, Rivadulla}  The above picture has been proposed for magnetic systems, but similar phenomena have been seen in non-magnetic systems as well.\cite{Morin,Granados_1}  This raises the possibility that a spin glass like model may not be the most appropriate to describe the dynamics of  phase separation in general. There has been other attempts to describe these phenomena in CMR manganites in terms of a phenomenological model where it has been proposed that the phase boundaries relax through a hierarchy of energy barriers.\cite{Levy, Ghivelder}  Though this is a commendable attempt we feel that this model does not really capture the essence of the phenomena because it does not consider what happens to the system while heating. There has been a more recent approach to understand the dynamics of phase separated systems based on the concept of kinetic arrest of a supercooled system into a magnetic glass and its subsequent de-arrest on warming.\cite{Chaddah10,Chaddah11}  But there are systems which are non-magnetic and  non-glassy and  it would be very interesting to make an attempt to find out the physics behind  the slow dynamics in such  systems because what it might reveal may turn out to have wider implications.\cite{Morin, Smith, Abe}

The perovskite oxide NdNiO$_3$, which undergoes a temperature driven M-I transition and has an associated phase separated state,\cite{Granados} we believe, can be taken as a model system to probe the dynamics of coexisting phases in non-magnetic, non-glassy systems. NdNiO$_3$ is a clean system which does not require any doping to see the  phase separated state and it is essentially unaffected by the application of a magnetic field and it does not show any evidence of glass transition.\cite{Mallik, Deven}  Moreover the phase separated state exists over a relatively large temperature range which makes it easy to study  slow dynamics and phase separation.

NdNiO$_3$ is a member of the series of compounds known as rare earth nickelates (RNiO$_{3}$), which are one of the few families of perovskite oxides that undergo a first order metal-insulator phase transition.\cite{Medarde}  These compounds crystallize in an orthorhombically distorted perovskite structure with space group P$_{bnm}$. The ground state of the nickelates (R$\neq$La) is insulating,\cite{Medarde,P.Lcorre} charge ordered\cite{Alonso_1,Alonso_2,SKim,Mizokawa,Staub} and antiferromagnetic.\cite{Torrance}  On increasing temperature from absolute zero these compounds undergo a temperature driven antiferromagnetic to paramagnetic transition, and an insulator to metal transition. Below the M-I transition temperature ($T_{MI}$), transport properties of NdNiO$_3$ exhibit a large hysteresis and in this region, the physical state of the  system is phase separated and it has been shown to consist of paramagnetic metallic and antiferromagnetic insulating phases.\cite{Granados, Medarde}

   In this work we use careful time and temperature dependent resistivity measurements on  NdNiO$_3$ to gain an understanding of its phase separated state and its dynamics. The phase separated state comes into existence on cooling the material below its M-I transition temperature and vanishes at sufficiently low temperature and does not form at all during subsequent heating.   We have provided a credible physical explanation along with a Monte Carlo simulation to describe all the behavior exhibited by our system. The simulation results are found to reproduce the experimental results quite well. Our results indicate that in the phase separated state of NdNiO$_3$, the metallic phase is present in its supercooled state. The supercooled metallic phase is metastable and it relaxes to the insulating equilibrium state giving rise to the various observed phenomena.

\section{EXPERIMENTAL DETAILS}

Polycrystalline NdNiO$_{3}$ samples in the form of 6\,mm diameter and 1\,mm thick pellets were prepared and characterized as described elsewhere.\citep{Massa}  The preparation method uses a high temperature of 1000$^{\circ}$C and a high oxygen pressure of 200\,bar.

All the temperature and time dependent measurements were done in a home made cryostat. To avoid thermal gradients in the sample during measurement it was mounted inside a thick-walled copper enclosure so that during the measurement the sample temperature would be uniform. It was found that mounting the sample in this fashion improved the reproducibility of the time dependence measurements significantly. A Lakeshore Cryotronics temperature controller was used to control the temperature and the temperature stability was found to be better than 3\,mK during constant temperature measurements.

Below $T_{MI}$ ($\approx 200$\,K) NdNiO$_{3}$ is not in thermodynamic equilibrium and slowly relaxes because of which the experimental data that we get depend on the procedure used for the measurement. The procedure we used was as follows. While cooling we start from 300\,K, and then record the data in steps of 1\,K interval after allowing the temperature to stabilize at each point. In between two temperature points the sample was cooled at a fixed cooling rate of 2\,K/min. After the cooling run is over we wait for one hour at 82\,K and then the heating data was collected at every one degree interval. The heating rate between temperature points was the same as the cooling rate used earlier. This cycle of measurements was repeated with a different cooling and heating rate of 0.2\,K/min also.

It was observed that the resistivity above 200\,K and below 115\,K does not show any time dependence, and it is also independent of measurement history. Thus to avoid the effect of any previous measurements, all time dependent experiments in the cooling run were done as follows: first take the sample to 220\,K (sufficiently above 200\,K), wait for half an hour, then cool at 2.0\,K/min to the temperature of interest and once the temperature has stabilized record the resistance as a function of time. In the heating run the time dependent resistivity was done in a similar fashion: first take the sample to 220\,K, wait for half an hour, then cool at 2.0\,K/min to 85\,K, wait for one hour, and then heat at 2.0\,K/min to the temperature of interest and once the temperature has stabilized record the resistance as a function of time.

The four probe van der Pauw method was used to measure the resistivity and standard precautions, such as current reversal to take care of stray emfs, were taken during the measurement. We also took care to ensure that the measuring current was not heating up the sample.

\section{RESULTS}

\begin{figure}[b]
\begin{centering}
\includegraphics[width=1\columnwidth]{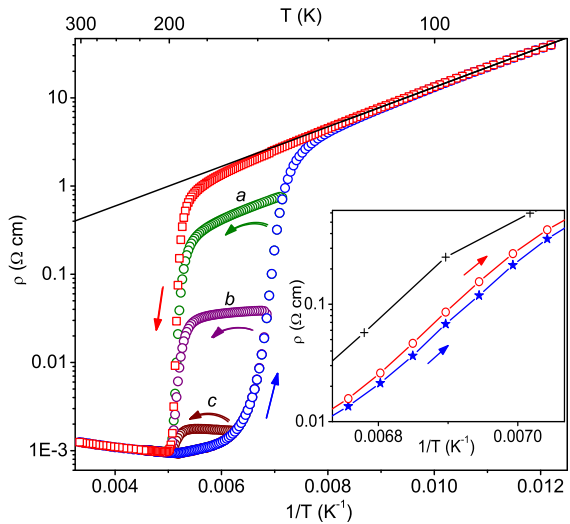}
\par\end{centering}

\caption{(Color online)
$\rho$ vs. $1/T$ plot for NdNiO$_{3}$. The blue 
circles represent cooling data and the red squares stand for heating
data (cooling/heating rate 2.0 K/min). The solid line is a least square
fit to the band gap insulator model below 115\,K. The curves labeled
\emph{a}, \emph{b} \& \emph{c} were taken as described in the text. The accuracy of the
data points is better than 1\% everywhere. The error does not exceed
0.8 m$\Omega$ cm anywhere. The inset shows $\rho$ vs $1/T$ for three
different cooling rates: lower curve (blue stars): 2\,K/min, middle
curve (red circles): 0.2\,K/min, upper curve (black pluses): infinitely
slowly (explained in text). The connecting lines are to guide the
eyes.}
\label{fig:R vs T}
\end{figure}

Figure \ref{fig:R vs T} shows the electrical resistivity of NdNiO$_{3}$ as a function of temperature. The resistivity is multiple valued, the cooling and heating data differing significantly from each other and forming a large hysteresis loop. The resistivity plot indicates that NdNiO$_{3}$ undergoes a relatively sharp M-I transition at about 200\,K while heating with a width of about 10\,K. In contrast, while cooling, the resistivity shows a rather broad M-I transition centered around 140\,K with a spread of about 40\,K. Below 115\,K or so, the heating and cooling data merge and the $\log\rho$ vs $1/T$ plot is linear. This indicates that the sample is insulating at low temperatures and, if the band gap is $\Delta$, the resistivity should follow the relation
\begin{equation}
\rho(T)=\rho_{0}e^{\frac{\Delta}{k_{B}T}}\label{eq:Band-gap-Resistivity}
\end{equation}
Below 115\,K both the heating and the cooling data fit quite well to this model, with a coefficient of determination, $R^{2}$= 0.99955. $\rho_{0}$ and $\Delta$ for the insulating region turn out to be 99\,m$\Omega$\,cm and 42\,meV respectively, which are in reasonable agreement with the values previously reported.\cite{Granados}

We collected more hysteresis data with different minimum temperatures such as 140\,K, 146\,K and 160\,K. In these measurements we cool the sample from 220\,K to one of the minimum temperatures mentioned
above and then heat it back to 220\,K, both operations being carried out at a fixed rate of 2\,K/min. Loops formed in this fashion are called minor loops and these are indicated by the labels \emph{a},
\emph{b} and \emph{c} in Figure \ref{fig:R vs T}. In the cooling cycle all the three minor loops coincide with the cooling curve of the full hysteresis loop. In the heating cycle, for loops \emph{a} and \emph{b }with lower minimum temperatures, the resistivity decreases with increasing temperature and joins the full loop at 200\,K. In the case of loop \emph{c,} as we increase the temperature, the resistivity
increases somewhat till about 188\,K and then it falls and joins  the full loop at 200\,K.

\begin{figure}[!t]
\begin{centering}
\includegraphics[width=1\columnwidth]{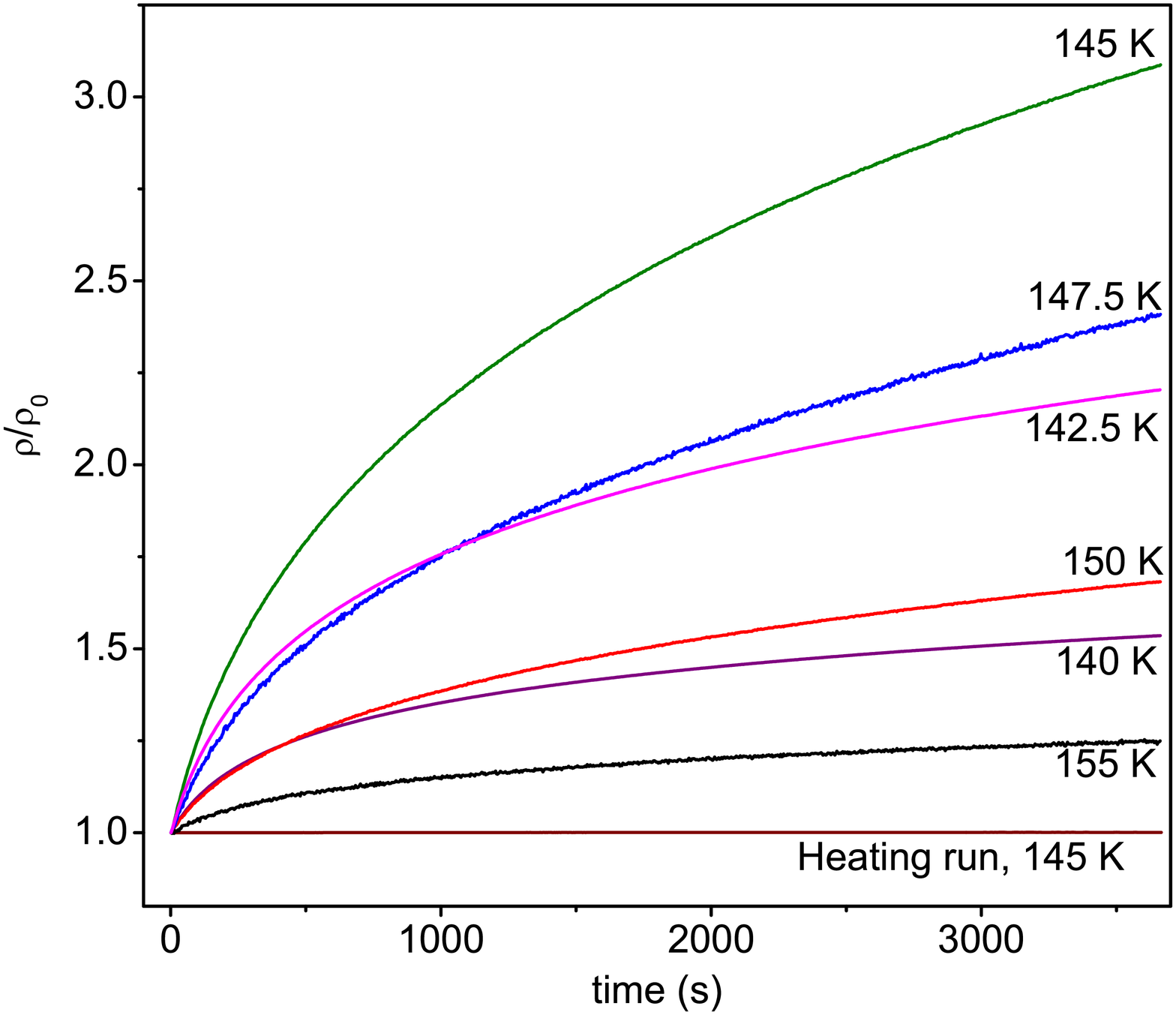}
\par\end{centering}

\caption{(Color online) Time dependence of resistivity while cooling, at various temperatures in the range 140 to 155\,K, for a period of one hour. The maximum time dependence is seen at 145\,K which is about 200\,\%. The curve at the bottom, which looks like a horizontal straight line, shows the increase in resistivity in a heating run taken at 145\,K and the total change for one hour in this case is less than 0.2\,\%. \label{fig:time-dependance} Not all the data are shown here to avoid clutter.}

\end{figure}

The resistivity also shows a noticeable dependence on the rate of temperature change in the cooling cycle as shown in the inset of Figure \ref{fig:R vs T}. The data for the lowest curve was collected at 2\,K/min and for the middle curve at 0.2\,K/min. The uppermost curve is an estimate obtained by extrapolating the time dependence data shown in Figure \ref{fig:time-dependance} to infinite time. We \emph{did not} see any dependence on rate of change of temperature when heating from 80\,K to 220\,K. This means that while cooling, below the M-I transition temperature (200\,K), the system is not in equilibrium and on the other hand while heating, the system is either in, or very close to, equilibrium. These observations are corroborated by the data shown in the next Figure.

A subset of the time dependent resistivity data taken while cooling is shown in Figure \ref{fig:time-dependance}. The data are presented as $\rho(t)/\rho(t=0)$ so that the values are normalized to unity at $t=0$ for easy comparison. We found that that below 160\,K, the resistivity of the sample increases with time considerably. A maximum relative increase in resistivity of about 200\,\% for a duration of one hour is seen at 145\,K, the time dependence being lower both above and below this temperature. We fitted the $\rho(T,t)$ curves in figure 2 to the stretched exponential function
\begin{equation}
\rho(t)=\rho_{0}+\rho_{1}\left(1-e^{-\left(\frac{t}{\tau}\right)^{\gamma}}\right)\label{eq:Stretched-Exponential}
\end{equation}

where $\rho_{0}$, $\rho_{1}$, $\tau$ and $\gamma$ are fit parameters. The fits are quite good with the $R^{2}$ value greater than 0.999 in most cases. See Table \ref{tab:Fit-parameters}. We note that the exponent $\gamma$ lies in the range $0.5<\gamma<0.6$ and $\tau$ has a peak around 147.5\,K. The variation of resistivity with time shows that the system slowly evolves towards an insulating state at a constant temperature. We collected data up to 12 hours (not shown here) to check whether the system reaches an equilibrium state, but found that it was continuing to relax even after such a long time.

 The magnitude of time dependence in heating runs (maximum $\approx0.2\%$) is negligible compared to what one gets in cooling runs (maximum$\approx200\%)$ (figure \ref{fig:time-dependance}), which suggest that in the heating run, below $T_{MI}$, the sample is almost fully insulating and stable.

\begin{table}
\begin{centering}
\begin{tabular}{|c|c|c|c|c|c|c|}
\hline
\# & T(K) & $\rho_{1}/\rho_{0}$ & $\tau$ ($10^{3}$s) & $\gamma$ & $\chi^{2}/DOF$ & $R^{2}$\tabularnewline
\hline
\hline
1 & 140.0 & 0.764(5) & 1.52(1) & 0.538(3) & 9.5 & 0.99971\tabularnewline
\hline
2 & 142.5 & 1.89(1) & 2.02(2) & 0.554(2) & 5.9 & 0.99981\tabularnewline
\hline
3 & 145.0 & 4.59(3) & 5.05(5) & 0.567(1) & 0.67 & 0.99993\tabularnewline
\hline
4 & 147.5 & 3.39(6) & 7.9(3) & 0.568(3) & 0.51 & 0.99978\tabularnewline
\hline
5 & 150.0 & 1.22(1) & 4.04(6) & 0.582(2) & 0.0014 & 0.99988\tabularnewline
\hline
6 & 155.0 & 0.391(5) & 2.6(1) & 0.557(7) & 0.0003 & 0.99845\tabularnewline
\hline
\end{tabular}
\par\end{centering}

\caption{\label{tab:Fit-parameters}Fit parameters for the time dependence data shown in Figure \ref{fig:time-dependance}. The degrees of freedom of the fits $DOF\approx1000$. The $\chi^{2}/DOF$ for 150\,K and 155\,K are too small, indicating that we have overestimated the error in resistivity in these cases. Anyway, we note that, the $R^{2}$ values are consistently good and indicate reasonably good fits.}
\end{table}

\begin{figure}[!t]
\begin{centering}
\includegraphics[width=1\columnwidth]{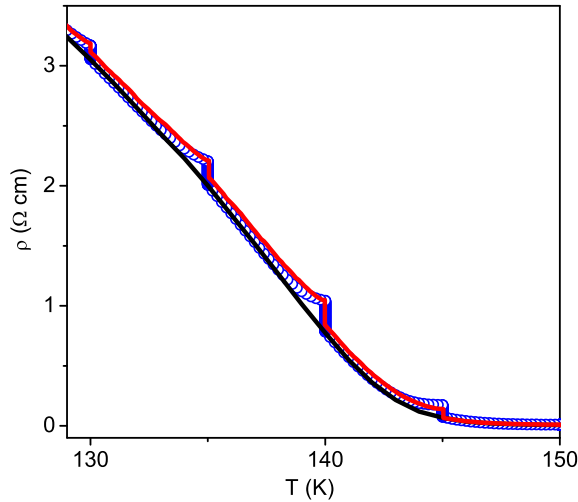}
\par\end{centering}

\caption{(Color online) Temperature dependence of $\rho$ on cooling with ageing by intermediate stops of one hour each (blue open circles) at 155, 150, 145, 140, 135, and 130 K. The black line show the resistivity without ageing. The red line shows the  results of the simulation discussed in section \ref{sec:simresults}.}

\label{fig:res-bifurcation}
\end{figure}

Figure \ref{fig:res-bifurcation} compares the resistivity in a cooling run with and without intermediate aging. These data were taken as follows: we start from 220\,K, come down to 160\,K at 2\,K/min, collect time dependence data for one hour and after that resume cooling at 2\,K/min and go down to 155\,K, collect resistivity time dependence again for one hour and so on down to 110\,K for every 5\,K interval. We note that when cooling is resumed after aging for an hour, $\rho(T)$ curve merges smoothly with the curve obtained without aging within a few kelvin. A rather similar observation has been reported in the phase-separated manganite La$_{0.5}$Ca$_{0.5}$Mn$_{0.95}$Fe$_{0.05}$O$_{3}$.\cite{Levy}

\section{DISCUSSION}

We have seen that, below $T_{MI}$, while cooling, the system is not in equilibrium and evolves with time which suggests that it is in a metastable state. Or, possibly, it consists of at least one metastable constituent. We infer that the resistivity of this metastable constituent slowly increases with time which probably implies that a slow metal to insulator transition is going on in the system. On the other hand while heating up from low temperature we saw that the system remains insulating all the way up to 188\,K and then it transforms to the metallic state by 200\,K. The resistivity was found to have negligible time dependence during the whole of the heating run. This result suggests that while heating from low temperature the system is in a stable equilibrium state.

It is well known that below a phase transition temperature a high temperature phase can survive as a metastable supercooled state. Based on this information we propose that below the M-I transition temperature, in the hysteresis region, our system consists of supercooled (SC) metallic regions and stable insulating regions. A supercooled metallic region would be separated from its stable insulating  state by an energy barrier which can be crossed with the help of perturbations such as thermal fluctuations, mechanical disturbances and so on. The crossing of the barrier from the metallic state into a stable insulating state can give rise to time dependence of resistivity in the system. Now, it is known that, on cooling a supercooled system it can either switch to a stable state or get kinetically arrested into a glassy state.  From the fact that no time dependence is observed while heating from low temperature towards the M-I transition temperature we rule out the possibility of any glassy phases forming in the system. For, if a glassy phase did form, time dependence would have been observed while heating from low temperature because of the de-arrest of the same. Consequently we are forced to conclude that the supercooled phases which are present while cooling would switch to the stable insulating state at a sufficiently low temperature. At a low enough temperature all of the material would be in the stable insulating state and hence subsequent heating will not show any time dependence.

   Based on the fact that no time dependence is observed above the M-I transition temperature, while heating, we rule out the possibility of the existence of superheated phases in our system.

   In the next subsections we get down to the nitty-gritty of the model we use to understand the experimental observations.

\subsection{The Model}

In a first order transition, a metastable SC phase can survive below the first order transition temperature ($T_{C}$), till a certain temperature called the limit of metastability ($T^{*}$) is reached.\cite{Chaikin,Chaddah,Chaddah1}  In the temperature range $T^{*}<T<T_{C}$ there is an energy barrier separating the SC metastable phase from the stable insulating state. The height of the energy barrier, $U$, can be written as $U\propto f(T-T^{*})$, where $f$ is a continuous and monotonic function of ($T-T^{*}$), and vanishes for $T\leq T^{*}$. As the temperature is lowered, at $T=T^{*}$, the SC metastable phase becomes unstable and switches over to the stable insulating state as the energy barrier becomes zero in this case.\cite{Chaikin} At $T>T^{*}$ the SC metastable phase can cross over to the stable insulating state with a probability (\emph{p}) which is governed by the Arrhenius equation
\begin{equation}
p\propto e^{-\frac{U}{k_{B}T}}\label{eq:Arrhenius}
\end{equation}
which tells us that the barrier will be crossed with an ensemble average time constant $\tau\propto1/p$. If we imagine an ensemble of such SC phases with the same barrier $U$, then the volume of the metastable phase will exponentially decay with a time constant $\tau$.

A transition from a supercooled state to a stable state is an avalanche-like transition which happens abruptly. Thus a single crystal or a crystallite, in the case of a polycrystalline material, would remain in the SC state above its temperature of metastability, and it will switch to the stable state as a whole when it is pushed over the free energy barrier by an energy fluctuation or if in the process of cooling  the temperature of metastability is attained. Examples for this would be the sharp M-I transitions along with hysteresis and time dependence observed in VO, V$_2$O$_3$, Li and Na.\cite{Morin,Smith,Abe}  It has been claimed that the presence of defects, strains and non-stoichiometry can give rise to different $T^*$'s for different crystallites, or it may even produce regions having different $T^*$'s inside a single crystal or crystallite.\cite{Chaddah, Imry, Soibel, Dagotto1}  We will call a crystallite or a region within a crystallite that will switch as a singe entity as a switchable region (SR) in our discussion.

Carrying forward the above arguments to our polycrystalline system, which is made up of tiny crystallites, we can say that it will be made up of a large collection of SR's. Each SR will have a unique  temperature of metastability, $T^*$, and a volume, $V$. The energy barrier $U$ that we discussed earlier, being an extensive quantity, will be proportional to the volume of the SR and we can write
\begin{equation}
U=Vf(T-T^{*})\label{eq:Barrier-height}
\end{equation}
 where $f$ is the continuous and monotonic function which vanishes for non-positive values of its argument. This means that the various SR's with their different $T^{*}$ and $V$ will have different energy barriers, which implies that the time constant $\tau$ will spread out and become a distribution of time constants depending on the distribution of the size and $T^*$ of the SR's. This can give rise to the volume of the metallic state decaying in a stretched exponential manner with time.\cite{Palmer,Ediger}   This behavior of the metallic volume with time will lead to the resistivity also evolving with time in a similar fashion. We shall see in the next subsection how the metallic volume and the resistivity are related to each other.

\begin{figure}[!t]
\begin{centering}
\includegraphics[width=1\columnwidth]{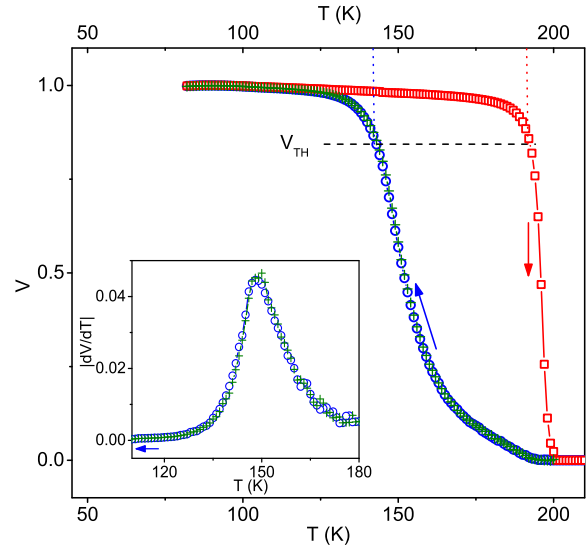}
\par\end{centering}

\caption{(Color online)Temperature variation of $V$ ($=1-f$), the insulating volume fraction. The blue circles show the cooling cycle (2\,K/min cooling rate) and the red squares represent the heating cycle. The green pluses represent the insulating volume fraction for 0.2\,K/min cooling rate. The dashed horizontal line with the label $V_{TH}(=$ 84\% insulating volume) represents the percolation threshold. The inset shows the variation of $|dV/dT|$ for the cooling cycle (blue circles: cooling rate 2\,K/min, and green pluses: cooling rate 0.2\,K/min).}

\label{fig:v-fraction}
\end{figure}


\subsection{Volume Fraction Calculation}

To check the validity of our model through numerical simulations we need to deal with the volumes of supercooled and stable phases. Our experimental measurements are of resistivity and we need to find a way of estimating the volumes of the different phases from our data. It is well known that the conductivity of a binary mixture consisting of insulating and metallic phases depends on their  respective volume fractions, geometries, distribution, and conductivities of the insulating and metallic phases($\sigma_{I}$ and $\sigma_{M}$).

We shall use an easy to use formula given by McLachlan,\cite{McLachlan}  which is based on a general effective medium (GEM) theory, for doing this conversion.  The McLachlan GEM equation has been successfully applied to a wide variety of isotropic, binary, macroscopic mixtures and it has been seen to work well even close to the percolation threshold.\cite{McLachlan,Hurvits,Kim}  If $\sigma_{E}$ is the effective electrical conductivity  of a binary MI mixture, the GEM equation says

\begin{equation}
(1-f)\frac{(\sigma_{I}^{1/t}-\sigma_{E}^{1/t})}{(\sigma_{I}^{1/t}+A\sigma_{E}^{1/t})}+f\frac{(\sigma_{M}^{1/t}-\sigma_{E}^{1/t})}{(\sigma_{M}^{1/t}+A\sigma_{E}^{1/t})}=0\label{eq:GEM}
\end{equation}
where $f$ is the volume fraction of the metallic phases and $A=(1-f_{c})/f_{c}$, $f_{c}$ being the volume fraction of metallic phases at the percolation threshold, and $t$ is a critical exponent which is close to 2 in three dimensions.\cite{Herrmann,Hurvits}  The constant $f_{c}$ depends on the lattice dimensionality, and for 3D its value is 0.16.\cite{Efros}

In order to calculate the volume fraction of metallic and insulating phases from Equation (\ref{eq:GEM}), we need their respective resistivities $\rho_{M}$ and $\rho_{I}$ as functions of temperature. $\rho_{M}(T)$ was obtained using $\rho_{M}=\rho_{0}+{\beta}T$, where $\beta$ is the temperature coefficient of resistivity estimated from the resistivity data above the M-I transition. $\rho_{I}(T)$ was calculated using Equation (\ref{eq:Band-gap-Resistivity}) with the parameters obtained by fitting the resistivity data below 115\,K. Using the above information and the resistivity data of the major loop shown in Figure \ref{fig:R vs T}, we calculated the volume fraction of metallic and insulating phases and it is shown in Figure \ref{fig:v-fraction}. In the cooling cycle the volume fraction of the insulating phases ($V=1-f$) slowly increases on decreasing the temperature below $T_{MI}$, while in the heating cycle it remains nearly constant up to about 185\,K, and then drops to zero by 200\,K. As can be seen from Figure \ref{fig:v-fraction} the percolation threshold for the cooling runs occurs at around 144\,K. Below this temperature there will be no continuous metallic paths in the system. 

\begin{figure}[!t]
\begin{centering}
\includegraphics[width=1\columnwidth]{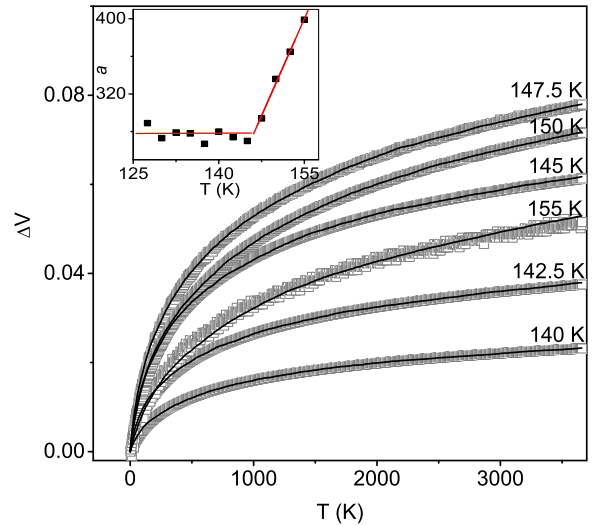}
\par\end{centering}

\caption{(Color online) $\Delta V$ vs time for 140\,K to 155\,K. The gray symbols show the values derived from experiments  and the black lines show the simulated values. The inset shows how the parameter $a$ depends on temperature. The red lines indicate the values we have chosen for the simulation.}
\label{fig:v-time}
\end{figure}

Figure \ref{fig:v-time} displays the increment  of insulating volume fraction, $\Delta V(exp)$, as a function of time that has been extracted from the time dependent resistivity data of Figure \ref{fig:time-dependance}. We see that the maximum time dependence in volume fraction is seen at 147.5\,K which is not the same temperature at which the maximum time dependence in resistivity is seen. It is interesting to note that the rate of change of metallic volume fraction, $|dV/dT|$, has a maximum around 147.5 K in the cooling runs (inset of Figure \ref{fig:v-fraction}) and it coincides with the temperature of the maximum volume time dependence. 

During cooling $|dV/dT|$ of the inset of Figure \ref{fig:v-fraction} represents the amount of volume that will change from the supercooled metallic to the insulating state for a unit temperature change. Now a small change in temperature from $T$ to $T -\delta T$ will result in all the supercooled SR's with their $T^{*}$ falling in that temperature range switching to the insulating state from the supercooled state. This means that $|dV/dT|\delta T$ at $T$ is a good measure of the volume fraction of supercooled metastable regions which have their $T^{*}$ close to $T$ within a temperature range $\delta T$. Thus $|dV/dT|$ of the cooling curve represents the volume distribution of $T^{*}$'s in the system. In coming to the above conclusion we have disregarded the small fraction of SR's that would be switching due to the time elapsed in covering the small temperature change. The justification for this is the fact that the $|dV/dT|$ values calculated from the 2\,K/min and 0.2\,K/min cooling curves are practically indistinguishable as can be seen from the inset of Figure \ref{fig:v-fraction}.

\subsection{Details of the Simulation}
We carried out Monte Carlo simulations to try to understand the experimental data. We take a distribution of $10^5$ SR's which have their volumes $(V)$ uniformly distributed in a certain range $(V_0, V_{max})$. These SR's were assigned a $T^*$  in such a way that the volume distribution of the supercooled metallic SR's matches with the $|dV/dT|$ curve shown in the inset of Figure \ref{fig:v-fraction}.  These metallic SR's have an energy  barrier given by Equation (\ref{eq:Barrier-height}) that separates them from the insulating equilibrium state. Now we assume that $f(T-T^*)$ has the simple power law form $f=c(T-T^*)^\alpha$ for $T>T^*$, where $c$ and $\alpha$ are positive constants, and $f=0$ for $T \leq T^*$. Using Equation \ref{eq:Arrhenius}, at any given temperature, the probability of switching from the metallic to the insulating state in a certain time interval is taken to be 

\begin{equation}
p=e^{-aV\frac{f(T-T^{*})}{T}}\label{eq:sim}
\end{equation}
with $a=c/k_B$.\cite{note2}

 Now to carry forward the simulation we have to estimate the parameters $V_0$, $V_{max}$, $a$, and $\alpha$. We decided to first simulate the time dependence of the insulating volume evolution shown in Figure \ref{fig:v-time} for this purpose. The time dependence of the insulating volume is generated as follows. To begin with we quench the simulated sample from above $T_{MI}$ to the temperature of interest. In this process all the SR's which have their $T^*$'s above the temperature of interest are switched from the metallic to the insulating state. Now the switching probability of each of the remaining metallic SR's is compared with a uniform random number in the interval [0,1) and if the switching probability is greater than the random number, the metallic SR is flipped to the insulating state and the increase in the insulating volume fraction is recorded. This operation is carried out for each one of the metallic SR's.  The time required for this process is taken to be the time interval at which the experimental data was recorded, which in our case is 3.6 seconds. We repeat this process one thousand ($n$) times  to generate data for one hour which is the time over which the experimental data was collected.  The parameter space of $V_0$, $V_{max}$, $a$, and $\alpha$ was searched to minimize the   error function

\begin{equation}
 Err (V_0, V_{max}, a, \alpha) = \sum_{i=0}^{n} \mid (\Delta V(exp)_i-\Delta V(cal)_i) \mid  \label{eq:error}
\end{equation}
where $\Delta V(exp)_i$ is the increase in insulating volume  fraction obtained from experimental data,    $\Delta V(cal)_i$ is the increase in insulating volume fraction calculated from the simulation and the subscript $i$ stands for the number of the time step. We found that as far as the volume of the SR's are concerned it is the ratio of $V_{max}$ to $V_0$ that decides the minimum value of $Err (V_0, V_{max}, a, \alpha)$ and not the individual values. So we fixed $V_0$ to be equal to unity and found that overall $V_{max} \approx 6$ and  $\alpha \approx 0.25$ gave the best values for the error function with parameter $a$ lying in the range 270 to 400 depending on the temperature. We refined the simulation by fixing the values of $V_{max}$ to be 6 and  $\alpha$ to be 0.25 and varying only $a$. We compare in Figure \ref{fig:v-time} the experimental data (gray symbols) and the simulation (black lines) and see that the simulation reproduces the experimental data quite closely. In the inset of the figure we show how the parameter $a$ varies with temperature.\cite{note1}  For further calculations we fixed the value of $a$ to be 280 up to 146\,K and above that temperature $a$ was taken to vary linearly with temperature as $a=280+14(T-146)$ and this is shown as red lines in the inset of Figure \ref{fig:v-time}.

It would have been ideal if all the parameters of the simulation turned out to be constants. But we find that we have to introduce some temperature dependence in $a$ to simulate the experimental results accurately.  The temperature dependence of $a$ could possibly be related to the percolation threshold which is at 144\,K. Below this temperature the background matrix for most of the SR's would be insulating while above this temperature the background  would gradually change to a metallic one as one goes away from the percolation threshold. We also note that an SR in its  insulating state  would have a slightly more distorted crystal structure along with a somewhat larger volume. Thus we see that the surroundings of an SR would be very different above and below 144\,K and this may have a bearing on the energy barrier the SR has to cross and this could result in a temperature dependent $a$.

In the simulation we quench the sample from above $T_{MI}$ to the temperature of interest while in the experiment the sample is cooled at a fixed cooling rate (2\,K/min) to the temperature of interest. In the simulation we can cool the sample slowly only if we know the parameters, and since we have no idea of the parameters to begin with, as a way out we decided to quench the sample. As we shall see soon, the simulation and  the experimental data agree quite well for the other experiments and hence it was felt that further refining of the simulation parameters would not improve the simulation very much and hence we decided to stick with the parameters obtained by quenching.

\subsection{Simulation Results}
\label{sec:simresults}

We will now describe the simulation results on resistivity hysteresis and minor loops, cooling rate dependence and rejuvenation.
\subsubsection{Hysteresis and Minor Loops in Resistivity}
    Hysteresis and minor loops in resistivity are calculated by first simulating how the insulating volume evolves as one cools and heats the (simulated) sample. The cooling was done by lowering the temperature by 0.12\,K first, during which process all those SR's with their $T^*$ in that 0.12\,K range are switched to the insulating state, followed by a time step of 3.6 seconds, during which all the remaining metallic SR's are given a chance to relax to the insulating state. We consider the above procedure to be a good approximation for cooling at a rate of 2\,K/min. This process is repeated as many times as required to complete the cooling part of the simulation. In the heating part of the simulation we again take temperature steps and time steps as before. But it may be noted that a temperature step while heating will not switch any of the SR's to the insulating state because all those SR's with $T^*$'s falling inside the temperature step would have already been switched to the insulating state during the preceding cooling process. During a time step all of the metallic SR's do get a chance to switch their states, but we point out that during heating the chance of a metallic SR switching to the insulating state would decrease drastically with increasing temperature because the energy barrier is a monotonically increasing function of $T-T^*$ and all the remaining metallic SR's would have their $T^*$ below the turning point, \emph{i.e.}, the temperature at which the cooling was stopped and heating started. After the volume simulation is completed we translate the results back to resistivity using Equation (\ref{eq:GEM}).

\begin{figure}[!t]
\begin{centering}
\includegraphics[width=1\columnwidth]{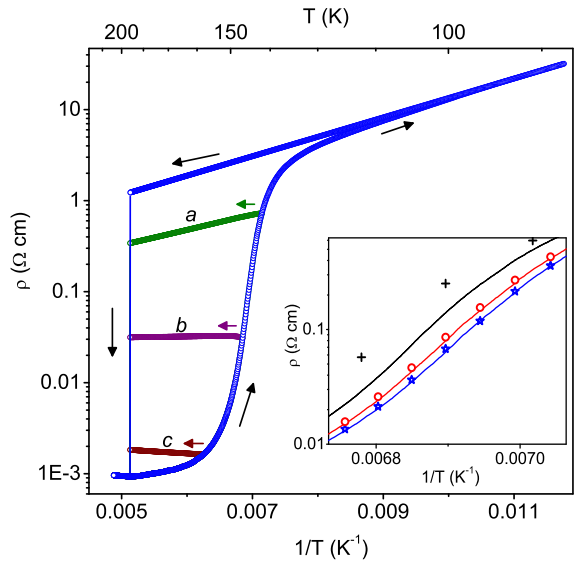}
\par\end{centering}

\caption{(Color online) Simulated results for the hysteresis along with minor loops. The labeling and color coding are the same as in Figure\,\ref{fig:R vs T}. Cooling rate dependence of resistivity is shown in the inset where the symbols represent the experimental data and the lines are from simulation. Blue star and blue line: 2\,K/min, red circles and red line: 0.2\,K/min, black line: 1\,K/hour simulation, black pluses: infinitely slow cooling rate, estimated from experiment.}
\label{fig:sim-hyst-coolrate}
\end{figure}

    In Figure\,\ref{fig:sim-hyst-coolrate} we show the simulated results for hysteresis and minor loops and we note that they are very similar to the experimental results shown in Figure\,\ref{fig:R vs T} except for the sharp transition in the simulation during heating. This can be attributed to the fact that we have not taken into account the broadening of the M-I transition due to disorder and finite size effects in our simulation.   It is interesting to note that the simulated minor loops reproduce quite closely some of the features of the experimental loops such as the positive slope of the heating (top) part of loop $a$ and the negative slope at the beginning of the heating part of loop $c$. The heating part of curve $a$ has a positive slope because in this case most of the SR's are in their insulating state and the temperature dependence of their resistivity is the dominating factor which determines its behavior. On the other hand the beginning of the heating part of curve $c$ has a negative slope because in this case most of SR's are in the metastable metallic state and the increase of their resistivity with time is the dominating factor here.

\subsubsection{Cooling Rate Dependence}

Simulating the cooling rate dependence is straightforward. The time steps were fixed at 3.6 seconds and the temperature steps were adjusted in size to simulate faster or slower cooling. The results are shown in the inset of Figure\,\ref{fig:sim-hyst-coolrate}. It is clear that the experimental data and the simulation agree very well for 0.2\,K/min and 2\,K/min cooling rates. We have also shown the simulation result for 1K\,/hour cooling rate and it is seen to fall in between the 2\,K/min data and the infinitely slow data. It is easy to see that a slower cooling rate will result in a higher resistivity because more of the SR's switch to the insulating state during slow cooling thus resulting in a higher resistivity.

\subsubsection{Intermediate Ageing}

Intermediate ageing is also fairly easy to simulate. We cool the simulated sample at 2\,K/min to the temperature of interest, do one thousand time steps to simulate waiting for one hour, and resume cooling and so on. In the inset of Figure \ref{fig:res-bifurcation} we have shown the simulation results for intermediate ageing and we see that the agreement between data and simulation is reasonable.

From Figure \ref{fig:res-bifurcation} it is clear that on resuming cooling after an intermediate stop of one hour the cooling curve merges with the curve obtained without ageing within about 3\,K or less. When we stop the cooling and age the sample at a fixed temperature the supercooled metallic SR's with relatively small $U$ will switch over to the insulating state. As can be inferred from Equation (\ref{eq:Barrier-height}) those SR's will have a small $U$ which have (i) their metastability temperature ($T^{*}$) close to their temperature ($T$) (ii) a small size. The major contributions to resistivity change will come from the relatively larger SR's making the transition from the metallic state to the insulating state. In the light of this the merger of the cooling curves with and without ageing within a small temperature change of about 3\,K means that most of the larger SR's which undergo the transition from metal to insulator during ageing have their $T^{*}$ within a few kelvin of the temperature of the sample.

\subsection{Predictive Power of the Model}


\begin{figure}[!t]
\begin{centering}
\includegraphics[width=1\columnwidth]{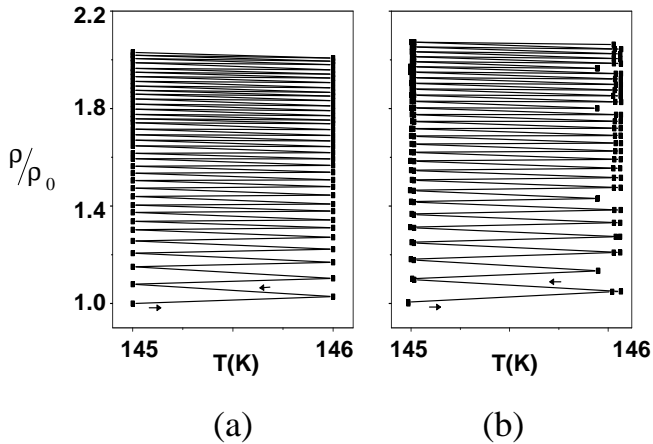}
\par\end{centering}

\caption{(a) shows the simulation results of how the resistivity evolves if the temperature of the sample is oscillated between 145\,K and 146\,K for one hour after cooling from above $T_{MI}$. $\rho_0$ is the resistivity of the sample when it reaches 145\,K after cooling from 220\,K at 2\,K/min. The arrows indicate that the resistivity keeps on increasing on thermal cycling. (b) shows the experimental results for the same.}

\label{fig:shaking}
\end{figure}

A model is no good if it cannot predict new results. So to test our model we decided to do a simulation first and then see whether experiment will reproduce it. In the simulation we oscillate the temperature of the sample between 145\,K and 146\,K for one hour immediately after cooling it from 220\,K to 145\,K. All temperature variations were done at 2\,K/min. Figure\,\ref{fig:shaking}(a) shows the results of the simulation. We see that the resistivity keeps on increasing on repeated thermal cycling; the increase per cycle is larger at the beginning and slowly it tapers off. We repeated the simulation on the heating cycle also, cooling the sample first from 220\,K down to 85\,K and then bringing it back up to 145\,K before starting the temperature oscillations. No detectable change in resistivity was seen with thermal cycling.

We expected that during the cooling cycle the temperature oscillations would disturb the sample and we would get a larger increase in resistivity than if the temperature were kept fixed at 145\,K. See Figure\,\ref{fig:time-dependance}. But instead, we find intriguingly, that the increase in resistivity  while oscillating the temperature is considerably less than what one gets at the constant temperature of 145\,K. Even more surprisingly it is less than that observed at 147.5\,K. It will be very interesting to see if experiment agrees with this prediction.

In Figure \ref{fig:shaking}(b) we show the experimental results obtained during thermal cycling after cooling the sample from 205\,K. One can see that the agreement between the simulation and the experiment is very good. We repeated the thermal cycling after cooling the sample down to 85\,K and bringing the temperature back up to 145\,K, just as in the simulation. Here also the experiment agreed very well with the simulation; no detectable change in resistivity was found.

The unexpectedly lower resistance rise when the temperature is oscillated after cooling can be attributed to (1) the smaller number of SR's available for switching because all those SR's with $T^*$ falling in the range 145\,K to 146\,K would have already switched to the stable insulating state the first time the temperature was lowered to 145\,K and (2) a higher average barrier height which an SR has to overcome, because the barrier height goes as $(T-T^*)^\alpha$.

\section{Conclusion}

Our experimental results and its excellent agreement with the simulation suggests strongly that while cooling the physical state of NdNiO$_{3}$ is phase separated below the M-I transition temperature; the phase separated state consists of SC metallic and insulating regions. A metastable metallic region switches from the metallic to the insulating state stochastically depending on the closeness of the temperature of metastability and the size of the region. At low temperature, below 115\,K or so, the system is insulating, all the SR's having switched over to the insulating state. While heating the SR's remain in the stable insulating state till the M-I transition temperature is reached and then switch over to the metallic state.

In our study we have developed a physical explanation of the  time dependence effects observed in  NdNiO$_3$ based on phase separation and supercooling. We believe that the physics developed here may have implications for first order solid-solid phase transitions in general. We have shown that our model has predictive power and we feel that it may serve as a useful template for understanding slow dynamics in hard condensed matter in such cases where there is phase separation. The apparently counter intuitive simulation prediction in the temperature oscillation experiment and its experimental confirmation gives us confidence that our model has captured the underlying physics of the problem truthfully.

\begin{acknowledgments}
DK thanks the University Grants Commission of India for financial support. KPR and DK thank Prof V. Subrahmaniam for useful discussions regarding the simulation. JAA and MJM-L acknowledge the Spanish Ministry of Education for funding the Project MAT2007-60536
\end{acknowledgments}


\begin{thebibliography}{30}
\bibitem{note2} The time interval referred to here is 3.6 seconds which is the time step used in the simulation. Please note that the parameters we would be extracting by minimizing the error function in Equation \ref{eq:error} not only depend on the sample properties but also on this time interval.
\bibitem{note1} With $a = 280$\,K$^{0.75}$/($\mu$m)$^3$ and taking $T-T^*$ to be 1\,K, the height of the energy barrier for a 1\,$\mu$m$^3$ sized SR turns out to be 24\,meV. At 145\,K this SR would flip, on an average, in 7 seconds  and this looks  a very reasonable number.
\end{thebibliography}
\end{document}